\documentclass[twocolumn,showpacs,preprintnumbers,amsmath,amssymb]{revtex4}

\usepackage{psfig}
\usepackage{graphicx}
\usepackage{dcolumn}
\usepackage{bm}

\begin{document}

\title{Spin-induced optical second harmonic generation \\
in the centrosymmetric magnetic semiconductors EuTe and EuSe}
\author{B. Kaminski$^{1}$, M. Lafrentz$^{1}$, R. V. Pisarev$^{2}$, D. R.
Yakovlev$^{1,2}$, V. V. Pavlov$^{2}$, V.~A.~Lukoshkin$^{2}$, A. B.
Henriques$^{3}$, G. Springholz$^{4}$, G. Bauer$^{4}$,
E.~Abramof$^{5}$, P. H. O. Rappl$^{5}$, and M.~Bayer$^{1}$}

\address{$^{1}$Experimentelle Physik 2, Technische Universit\"at Dortmund,
D-44221 Dortmund, Germany}
\address{$^{2}$ Ioffe Physical-Technical Institute, Russian
Academy of Sciences, 194021 St. Petersburg, Russia}
\address{$^{3}$ Instituto de F\'{\i}sica, Universidade de S\~ao Paulo,
05315-970 S\~ao Paulo, Brazil}
\address{$^{4}$ Institut f\"ur Halbleiter- und Festk\"orperphysik,
Johannes Kepler Universit\"at Linz, 4040 Linz, Austria}
\address{$^{5}$ LAS-INPE, 12227-010
S\~ao Jos\'e dos Campos, Brazil}

\date{\today}

\begin{abstract}
Spectroscopy of the centrosymmetric magnetic semiconductors EuTe and
EuSe reveals spin-induced optical second harmonic generation (SHG)
in the band gap vicinity at 2.1-2.4~eV. The magnetic field and
temperature dependence demonstrates that the SHG arises from the
bulk of the materials due to a novel type of nonlinear optical
susceptibility caused by the magnetic dipole contribution combined
with spontaneous or induced magnetization. This spin-induced
susceptibility opens access to a wide class of centrosymmetric
systems by harmonics generation spectroscopy.
\end{abstract}


\maketitle Nonlinear optics is a highly active field of basic and
applied research with optical harmonics generation playing a
particularly important role~\cite{Shen,Boyd}. Harmonics generation
is associated with higher order optical susceptibilities, and opens
access to unique information about the crystallographic, electronic
and magnetic structure of solids that is inaccessible by linear
optics~\cite{Shen,Boyd,Fiebig1}. Second harmonic generation (SHG)
has attracted most interest, because of its exceptional sensitivity
to space and time symmetry violations \cite{Fiebig1} and its
importance for  technological applications. Spectroscopy of
semiconductors using SHG has been, however, mostly limited to narrow
spectral ranges~\cite{Wagner,Bergfeld}. Recently, SHG was studied in
detail for the \textit{noncentrosymmetric} semiconductors GaAs, CdTe
and (Cd,Mn)Te ~\cite{Pavlov,Sanger2}, where SHG is allowed in
electric-dipole (ED) approximation. Two mechanisms of
magnetic-field-induced SHG have been disclosed, based on changing ED
contributions by mixing with magnetically-induced terms. In
\textit{centrosymmetric} materials with inversion symmetry SHG is
forbidden in ED approximation, which imposes severe restrictions on
the crystalline solids and artificial structures that can be
explored by SHG.

This restriction can be overcome by processes based on
magnetic-dipole (MD) or electric-quadrupole (EQ) nonlinear
susceptibilities. Other opportunities may be opened up by external
or internal perturbations that break either space-inversion or
time-reversal symmetry. For example, an applied electric field
breaks the space-inversion symmetry in centrosymmetric materials so
that ED-SHG becomes allowed~\cite{Terhune}. It would be highly
attractive to find the counterpart SHG related exclusively to MD
contributions triggered by applied magnetic fields or magnetic
ordering. Evidently, the search for such mechanisms is facilitated
in centrosymmetric materials, where crystallographic ED and EQ
contributions to SHG vanish.

In this Letter we report on spin-induced SHG in the centrosymmetric
magnetic semiconductors EuTe and EuSe. No SHG was detected in the
antiferromagnetic and paramagnetic phases. However, when a magnetic
field is applied, SHG arises due to breaking of the
antiferromagnetic order or by polarization of the paramagnetic
phase, both resulting in appearance of a net magnetization. The
observed spin-related nonlinearities arise due to a novel type of
nonlinear optical susceptibility caused by the MD contribution in
combination with spontaneous or induced magnetization.

Europium chalcogenides Eu\textit{X} (\textit{X}=O, S, Se, and Te)
are  magnetic semiconductors crystallizing in the centrosymmetric
cubic rock salt structure ${m3m}$. They possess unique physical
properties determined by the electronic structure in which the
strongly localized 4\textit{f}$^7$ electrons of Eu$^{2+}$ ions with
spin $S=7/2$ are involved~\cite{Wachter}. Eu\textit{X} are classical
Heisenberg magnets where the competition between nearest and
next-nearest neighbor exchange integrals results in magnetic phase
diagrams that can include antiferro- (AFM), ferri- (FIM), and
ferromagnetic (FM) ordering as well as a paramagnetic phase at
elevated temperatures~\cite{Wachter,Lechner}. EuTe is
antiferromagnetic  with a N\'eel temperature $T_N$=9.6~K and a
critical field $B_c=7.2$~T above which it becomes ferromagnetically
saturated. EuSe is metamagnetic with $T_N=4.6$~K and shows a mixed
AFM and FIM ordering below 2.8~K. At $T<2$~K and in a magnetic field
above a critical value of 0.2~T EuSe is in the FM
phase~\cite{Lechner}. Eu\textit{X} exhibit strong linear
magneto-optical effects~\cite{Schoenes,ABH1,ABH2}, due to what they
attract interest for potential applications in spin injection and
magneto-optical
devices~\cite{Schmehl,Springholz,Trbovic,Santos,Miao}.  Nonlinear
optical properties of these materials have not yet been explored.

The experimental technique is described in Ref.~\cite{Pavlov}. SHG
spectra were recorded in transmission geometry using 8~ns pulses
with a 10~Hz repetition rate generated by an optical parametric
oscillator. Experiments were performed on EuTe and EuSe layers grown
by molecular-beam epitaxy on (111)-oriented BaF$_2$
substrates~\cite{Heiss,ABH1}. The ~1~$\mu$m thick layers were capped
with a 40-nm-thick BaF$_2$ protective layer and the high sample
quality was confirmed by x-ray analysis. The sample temperature was
varied from 1.4 to 50~K. Magnetic fields up to $B$=10~T were applied
in the Voigt geometry.

\begin{figure}[ht]
\includegraphics[width=0.5\textwidth,keepaspectratio=true]{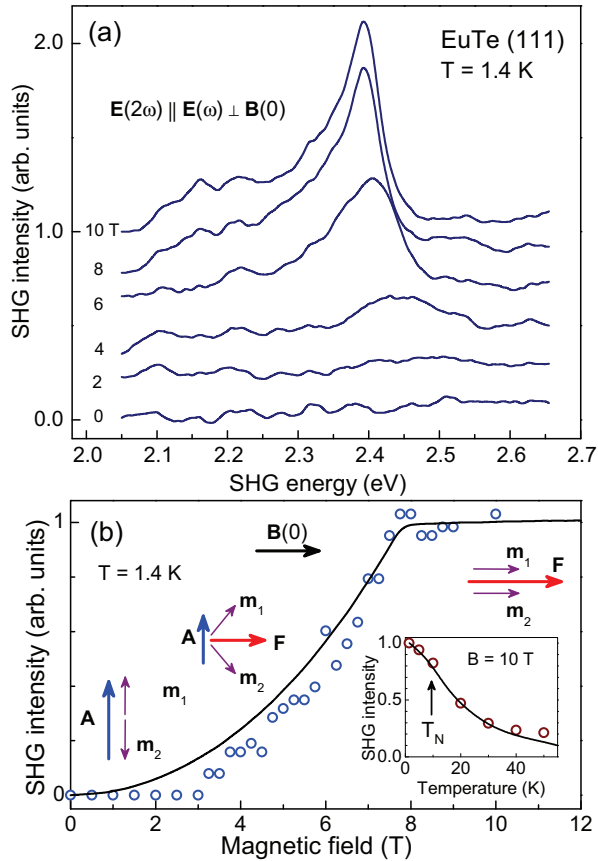}
\caption{(Color online) (a) SHG spectra of EuTe in different
magnetic fields. Spectra are offset by 0.2 relative to each other.
(b) Integral SHG intensity as function of magnetic field. Solid line
gives normalized magnetization $M^2(B)$~\cite{Oliveira}. Inset shows
temperature dependence of peak intensity at $B=10$~T. Line gives
normalized $M^2(T)$  after Ref.~\cite{Oliveira}.} \label{specfield}
\end{figure}
Figure~1(a) displays the SHG spectra of EuTe recorded at different
magnetic fields. At zero field, no SHG signal was detected in a wide
temperature range below and above $T_N$. However, SHG appears at
finite $B$ in the vicinity of the band gap and its structure with a
maximum at 2.4~eV and a shoulder at 2.2~eV is in good agreement with
EuTe absorption spectra~\cite{ABH1,ABH2}. As shown in Fig.~1(b), the
integrated SHG intensity increases with field and saturates for
$B>7.5$~T. Remarkably, it follows the square magnetization of EuTe
represented by the solid line,  which saturates above $B_c$=7.2~T
where all spins are collinearly aligned~\cite{Oliveira}. As is
indicated by the arrows in Fig.~1(b), this magnetization stems from
a continuous transformation of the AFM ordering at $B$=0 to the FM
one above $B_c$. At a fixed magnetic field of $B$=10~T, the SHG
signal continuously decreases with increasing temperature and
vanishes at about 50~K. As demonstrated by the insert of Fig.~1(b)
this decrease follows approximately the $M^2(T)$
dependence~\cite{Oliveira}. Above $T_N$ the SHG signal is obviously
related to the paramagnetic spin polarization.

It is evident from Fig.~1 that the SHG mechanism in EuTe is
controlled by the spin polarization of Eu$^{2+}$ induced by external
magnetic fields. The experiments on EuSe confirm this conclusion
despite its complicated magnetic phase diagram~\cite{Lechner}.
Figure 2(a) shows that in the vicinity of the optical band gap
around 2.1-2.4~eV, again a clear SHG signal appears at finite $B$,
which is absent at $B$=0. The magnetic field dependence of the
corresponding SHG intensity is shown in Fig.~2(b) for two
experimental geometries with
$\mathbf{E}(2\omega)\bot\mathbf{E}(\omega)$ and
$\mathbf{E}(2\omega)\|\mathbf{E}(\omega)$. In both cases, the SHG
intensity increases in a stepwise manner with increasing field, and
shows two saturation regions, one between 0.01 and 0.2~T and a
second one above 0.2~T. These steps are in good agreement with the
critical fields for the magnetic phase transitions of
EuSe~\cite{Lechner}. It gives us a clear proof that the measured SHG
arises from the bulk of the sample and not from the surface
\cite{Sipe}, because critical fields at the surface, and in
particular in antiferromagnets, radically differ from those in bulk.

\begin{figure}[ht]
\includegraphics[width=0.5\textwidth, keepaspectratio=true]{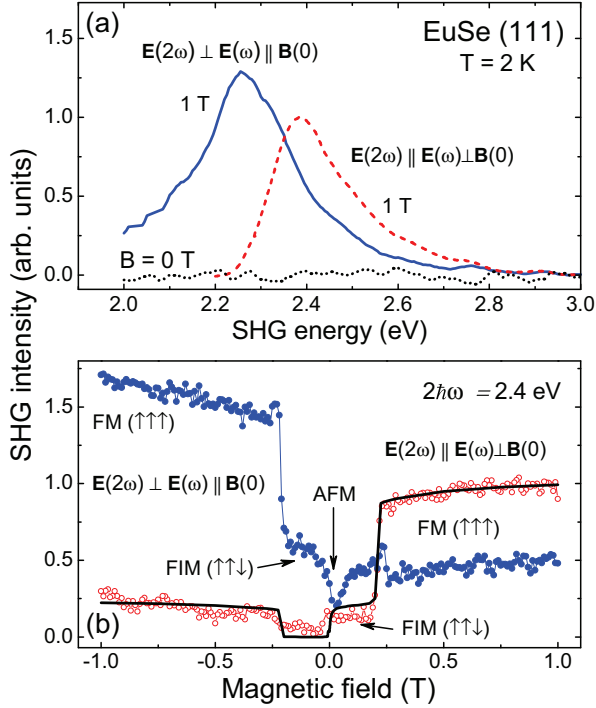}
\caption{(Color online) (a) Magnetic-field-induced SHG spectra in
EuSe shown for zero field and for a saturation field of +1~T for two
different measurement geometries. (b) SHG intensity vs magnetic
field. The line gives normalized $[a+bM(B)]^2$ with $b/a=4$ and
$M(B)$ after Ref.~\cite{Lechner}.} \label{FielddepEuse}
\end{figure}
Two contributions to the nonlinear optical polarization can be
expected in the Eu\textit{X} magnetic semiconductors. The one
related to the crystallographic magnetic dipole (CMD)
is~\cite{Pershan}
\begin{eqnarray}
P{^{CMD}_{i}}(2\omega)= i\varepsilon_0\chi^{(2)}_{ijk}E_j(\omega)B_k(\omega),
\end{eqnarray}
where $E_j(\omega)$ and $B_k(\omega)$ are the electric and magnetic
fields of the fundamental light wave, respectively.
$\chi^{(2)}_{ijk}$ is an axial third-rank tensor allowed in any
medium~\cite{Birss}. The same type of tensor describes the Faraday
effect, if in Eq.~(1) the light field $B_k(\omega)$ is replaced by
an external magnetic field $B_k(0)$, and the polarization
$P_{i}(\omega)$ is excited at the fundamental frequency. In view of
the strong Faraday effect in Eu\textit{X} compounds achieving
$~10^6$~deg/cm~\cite{Schoenes} a significant MD contribution to the
SHG is expected~\cite{Pershan}. In bulk Eu\textit{X} with point
group $m3m$ the tensor $\chi^{(2)}_{ijk}$ has only one non-vanishing
independent component $xyz(c3)$=-$xzy(c3)$~\cite{Birss}, which,
however, does not lead to any SHG intensity
$I(2\omega)$$\propto$$|\mathbf{P}^{CMD}(2\omega)|^2$ because of the
alternating sign change for every axis permutation.

Layers of EuTe and EuSe grown on BaF$_2$ substrate are known to
acquire a small mismatch between sample and substrate lattice
constants resulting in a weak trigonal distortion along the
$[111]$-axis. Evidently this causes a symmetry reduction of the thin
EuTe and EuSe layers to the trigonal centrosymmetric point group
$\overline{3}m$ in the proximity of the interface. The tensor
$\chi^{(2)}_{ijk}$ in this point group has one independent component
$xyz(6)$~\cite{Birss} which may produce a small crystallographic MD
contribution to the SHG signal.

A new type of nonlinear polarization can be induced if the parent
crystal symmetry is broken by either magnetic field or magnetic
ordering, both of which we introduce with the magnetic parameter
$\textbf{M}(0)$. The induced magnetic dipole (IMD) nonlinear
polarization is
\begin{eqnarray}
P{^{IMD}_{i}}(2\omega)=
\varepsilon_0\chi^{(3)}_{ijkl}E_j(\omega)B_k(\omega)M_l(0),
\end{eqnarray}
where $\chi^{(3)}_{ijkl}$ is a polar forth-rank tensor~\cite{Birss}.

In EuTe the magnetic ordering below $T_N$ can be characterized by
the magnetic moments \textbf{m}$_1$ and \textbf{m}$_2$ of the two
sublattices with $|m_1|=|m_2|$. To describe the magnetic behavior
of an antiferromagnet in external magnetic field we introduce a FM
vector \textbf{F}=\textbf{m}${_1}$+\textbf{m}${_2}$ and an AFM
vector \textbf{A}=\textbf{m}${_1}$-\textbf{m}${_2}$. The scheme in
Fig.~1(b) shows the sublattice reorientation  when the magnetic
field is increased. Though \textbf{F} and \textbf{A} are composed
of the same vectors \textbf{m}${_1}$ and \textbf{m}${_2}$, their
transformation properties are different. \textbf{F} changes sign
under time reversal, but not under space inversion, and thus
transforms as a MD. The AFM vector \textbf{A} does not induce any
SHG signal, which qualitatively can be understood as follows: at
$B$=0 \textbf{m}$_1$ and \textbf{m}$_2$ are oriented
antiferromagnetically. Each magnetic sublattice induces a SHG
signal via MD according to Eq.~(2) with
$\mathbf{M}(0)=\mathbf{m}_{1,2}$, but destructive interference
from oppositely oriented sublattices annihilates the SHG signal
since the relevant nonlinear polarization is an odd function of
\textbf{M}(0). With increasing magnetic field the AFM ordering is
transformed into a FM one and the destructive interference is
continuously reduced. In this case \textbf{M}(0) in Eq.~(2) should
be associated with the ferromagnetic vector \textbf{F}. The SHG
signal increases with magnetic field and reaches saturation when
the two sublattices become oriented ferromagnetically. Above $T_N$
in the paramagnetic phase $\mathbf{M}(0)=\chi_p\textbf{B}$, where
$\chi_p$ is the paramagnetic susceptibility.

The rotational anisotropy of the SHG intensity detected for
simultaneous rotation of linear polarizers for the fundamental and
SHG light is a characteristic feature of the coherent SHG process
giving an in-depth view on the symmetries involved. Corresponding
diagrams for EuTe and EuSe are shown in Fig.~3 for the parallel
[$\mathbf{E}(2\omega)\parallel \mathbf{E}(\omega)$] and
perpendicular [$\mathbf{E}(2\omega)\perp \mathbf{E}(\omega)$]
configuration. For EuTe the rotational anisotropies are twofold.
In EuSe the rotational anisotropy is twofold in the parallel
configuration but is transformed into a distorted fourfold
anisotropy for the perpendicular case.

To model the rotational anisotropy of SHG intensity, the
interference between IMD [Eq.~(1)] and CMD [Eq.~(2)] contributions
should be taken into account
\begin{eqnarray}
I(2\omega)\propto |\mathbf{P}^{IMD}|^2 + |\mathbf{P}^{CMD}|^2 \pm
2|\mathbf{P}^{IMD}\mathbf{P}^{CMD}|.
\end{eqnarray}
Here the signs $\pm$ correspond to opposite orientations of
$\textbf{M}(0)$. The SHG intensities for the parallel and
perpendicular configurations of fundamental and SHG light
polarizations are
\begin{eqnarray}
\begin{split}
I_{\parallel}(2\omega)\propto&
[\pm\frac{F}{6}(\chi_{xxxx}+5\chi_{xxyy}-\chi_{xyxy}-\chi_{xyyx})\cos{\varphi}-\\
&\chi_{xyz}\cos{3(\alpha + \varphi)}]^2, \end{split}\\
\begin{split}
I_{\perp}(2\omega)\propto&
[\pm\frac{F}{6}(\chi_{xxxx}-\chi_{xxyy}+5\chi_{xyxy}-\chi_{xyyx})\sin{\varphi}+\\
&\chi_{xyz}\sin{3(\alpha+\varphi)}]^2, \end{split}
\end{eqnarray}
where $\varphi$ is the angle between the polarization plane of the
fundamental light and the crystallographic $[11\overline{2}]$ axis
and $\alpha$ is the sample azimuthal angle. As shown by Figs.~3(e)
and 3(f), the IMD contribution calculated for the point group $m3m$
results in a twofold diagram, whereas the CMD contribution for the
point group $\overline{3}m$ results in a sixfold diagram. The
modeling with the IMD contribution only, shown by solid lines in
Figs.~3(a) and (d), are in satisfactory agreement for EuTe. For EuSe
the parallel configuration is in qualitative agreement, but not the
fourfold anisotropy in perpendicular configuration. The modeling
with both contributions are shown by the shaded areas. Very good
agreement is achieved for both EuTe and EuSe. The fitting procedure
reveals a $P^{IMD}$/$P^{CMD}$ ratio of $\sim$7/1 for EuTe and
$\sim$4/1 for EuSe. Thus, the IMD contribution dominates over the
CMD one, which allows us to conclude that the MD-SHG process in
Eu$X$ is mainly determined by the spin-induced mechanism. This is
also confirmed by the vanishing SHG signal at elevated temperatures,
shown in the insert of Fig.~1(b).
\begin{figure}[bp]
\includegraphics[width=0.5\textwidth,keepaspectratio=true]{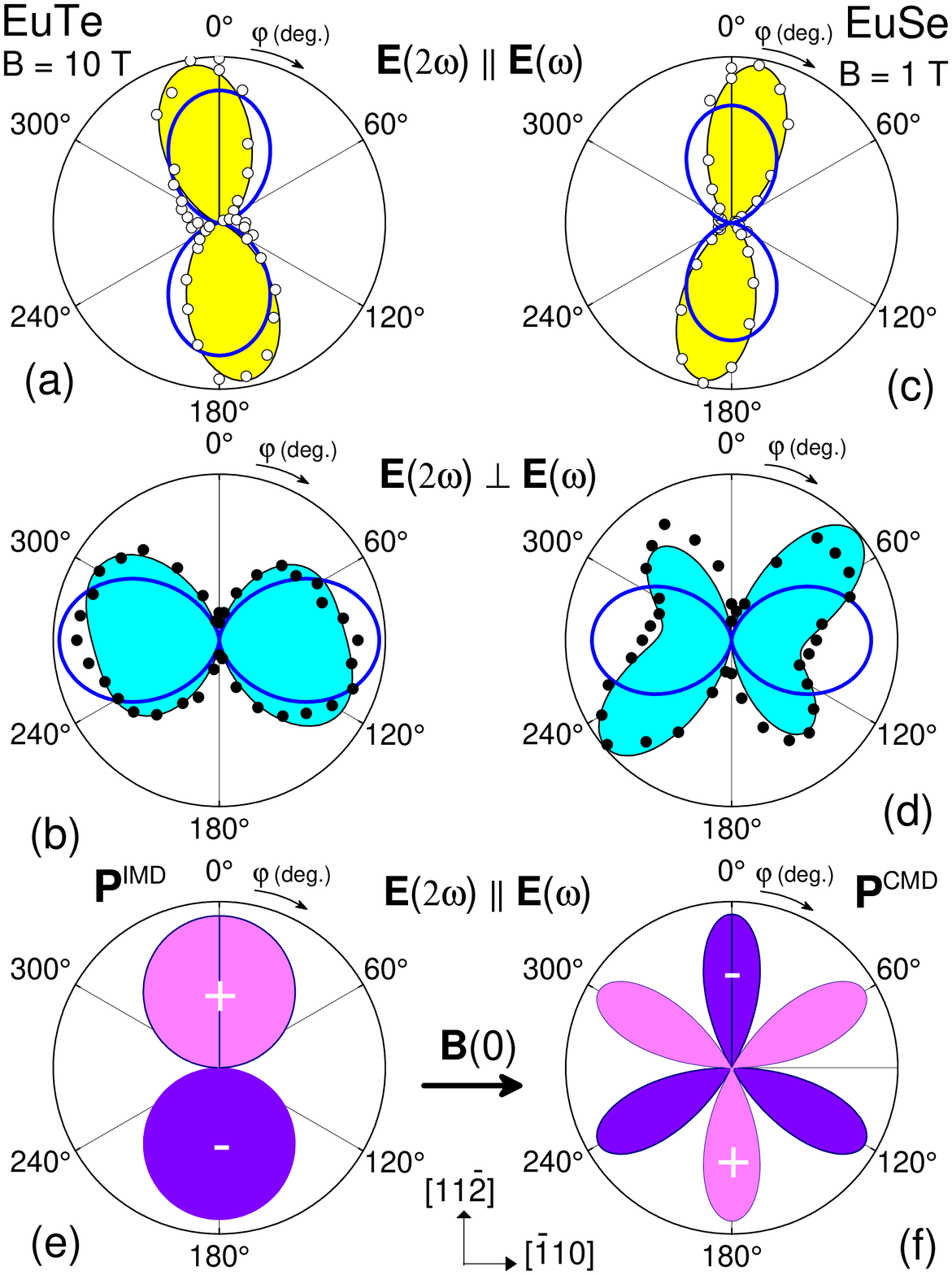}
\caption{(Color online) Polar plots of experimental SHG intensity
data (dots) in EuTe (a,b) and EuSe (c,d) measured at 2.4~eV. Best
fits based on Eqs.~(4) and (5), taking into account the IMD (IMD and
CMD), are shown by solid lines (shaded areas). Polar plots for
calculated IMD (e) and CMD (f) nonlinear polarizations with $m3m$
and $\overline{3}m$ symmetries, respectively.}
\label{Rotational_diagrams}
\end{figure}

The negligible role of the CMD contribution for EuTe is confirmed by
the fact that the magnetic field dependence of the SHG intensity
$I(2\omega)$$\propto$$M^2$, as seen in Fig.~1(b). To account for the
interference of CMD and IMD contributions in EuSe we compare in
Fig.~2(b) $I(2\omega)$ with $[a+bM(B)]^2$ dependence taking $b/a$=4
similar to $P^{IMD}/P^{CMD}$=4 for the saturated magnetization. The
observed asymmetry in the field dependence is well explained.
Therefore, the observed SHG is due to the FM component of the spin
system in Eu\textit{X}, which induces the MD contribution to SHG.
The role of the external magnetic field is to induce the
ferromagnetic component \textbf{F}. SHG signals are observed also
above $T_N$, when the magnetic field polarizes the Eu$^{2+}$ spins
in a paramagnetic phase. Thus, application of the magnetic field to
Eu\textit{X} leads to a new type of MD nonlinearity. It can be
treated as a counterpart to electric field application to
centrosymmetric media which breaks space inversion symmetry and
allows ED-SHG processes.

The spin-induced nonlinearities in Eu\textit{X} can be analyzed in
the framework of a microscopic model. For EuTe the valence band is
formed by the 5$p^6$ orbitals of Te$^{2-}$ and the conduction band
by the 5$d$ and 6$s$ orbitals of Eu$^{2+}$ ~\cite{Wachter}. The 5$d$
levels are split by the crystal field into $t_{2g}$ and $e_g$
subbands. The localized $4f^7$ states of Eu$^{2+}$ are close to the
top of the valence band. The optical band gap is determined by
$4f^7$$\rightarrow$$5d(t_{2g})$ transitions. The SHG is explained
taking into account nonlinearities involving an $^8S_{7/2}$ ground
state and electronic levels within the conduction band, given by the
$^7F_JY$, where $J=0,\ldots,6$ and $Y$ is one of the three orbital
states in the $5d(t_{2g})$ subset~\cite{ABH2}. The main
contributions to the spin-induced nonlinear susceptibility from each
magnetic sublattice is
\begin{eqnarray}
\frac{\langle^8S_{7/2}|x|^7F_JY\rangle\langle
^7F_JY|L_y|^7F_J'Y'\rangle\langle ^7F_J'Y'|x|^8S_{7/2}\rangle}{\left[E\left(^7F_J'Y'\right)-\hbar\omega\right]\left[E\left(^7F_JY\right)-2
\hbar\omega\right]},
\end{eqnarray}
where $L_y$ is the magnetic dipole operator and $x$ is position of
all seven electrons. It must be emphasized that only the magnetic
dipole operator can couple the $Y$ and $Y'$ states, whereas their ED
or EQ couplings are symmetry forbidden. Calculations, whose details
will be given elsewhere, lead to vanishing of the spin-induced
nonlinear susceptibility in the AFM phase, because the contributions
from both magnetic sublattices cancel each other. The susceptibility
increases when the Eu$^{2+}$ spins are tilted towards the magnetic
field. This explains the observed relation between the SHG intensity
and the sample magnetization.

In conclusion, spin-induced SHG is found in the centrosymmetric
magnetic semiconductors EuTe and EuSe. The established
magnetic-dipole mechanism induces bulk SHG polarizations either by
the ferromagnetic component of the magnetic structure, or by the
spin polarization in the paramagnetic phase. This new type of
spin-induced nonlinear susceptibility can appreciably increase the
number of centrosymmetric bulk materials, thin films, and artificial
structures accessible to nonlinear optics.

This work was supported by the Deutsche Forschungsgemeinschaft
(YA65/4-1), the Russian Foundation for Basic Research, Russian
Academy Program on Spintronics, FoNE of the European Science
Foundation, the Brazilian Agencies FAPESP and CNPq, the Austrian
Science Funds (I80-N20) and the FWF (Vienna, Austria).

\end{document}